\documentstyle[preprint,tighten,aps,psfig]{revtex}

\begin{document}
\preprint{TTP96-28, hep-ph/9608366}
\draft

\date{August 1996}
\title{Nonfactorizable QCD and Electroweak Corrections to the Hadronic
     \\  $Z$ Boson Decay Rate}
\author{Andrzej Czarnecki and Johann H. K\"uhn}
\address{Institut f\"ur Theoretische Teilchenphysik, 
Universit\"at Karlsruhe,\\
D-76128 Karlsruhe, Germany}
\maketitle

\begin{abstract}
We present an analysis of two-loop mixed QCD and electroweak
corrections to the decay of the $Z$ boson into light quarks. 
We find that the naive factorization of QCD and electroweak
corrections does not describe correctly the two-loop effects.
The nonfactorizable corrections shift the width of the $Z$ boson
by approximately $-0.55(3)$ MeV and increase the central value of the
strong coupling constant determined at LEP by 0.001.
\end{abstract}

\vspace*{2mm}
With a total number of about four million  hadronic $Z$ decays
collected at each of the four LEP experiments a final precision in the
$Z$ boson decay rate of about 2 MeV is expected.  At the same
time the ratio $R_{\rm had} = \Gamma(Z\to {hadrons})/\Gamma(Z\to
\mu^+\mu^-)$ will be measured with an accuracy of about 0.1 per cent.
These measurements will determine $\alpha_s$ with a remaining
uncertainty of about $\delta \alpha_s = \pm 3\times 10^{-3}$. This
puts not only severe constraints on the experimental analysis but also
on the theoretical understanding of subtle higher order effects.  QCD
corrections have been calculated up to order $\alpha_s^3$
\cite{alpha3} and leading as well as subleading electroweak
corrections have been calculated up to two loops \cite{ew}.

Another important class of effects is provided by the interplay between
electroweak interactions and QCD.  The relation between $M_W$, $M_Z$,
$G_F$, and $\alpha$ is affected by the large top quark mass and again
two and even three loop self-energies have been evaluated to arrive at
accurate predictions.  In addition there is of course the large class
of ``reducible'' corrections which originate from one-loop electroweak
diagrams, multiplied by the QCD correction factor $(1+\alpha_s/\pi)$.
The remaining two-loop effects are induced by nonfactorizable vertex
corrections.  With an order of magnitude of ${\alpha_{\rm
weak}\over\pi} {\alpha_s\over\pi}\approx 0.4\times 10^{-3}$ and an
unknown coefficient, they are not {\em a priori} negligible for an
analysis at the 0.1 per cent level.  In particular it is evident that an
ansatz assuming factorization between QCD and weak effects cannot be
valid for the irreducible vertex diagrams. For the $Z$ decay rate into
bottom quarks this has been confirmed by the calculation of the
leading contributions which are enhanced by a factor $m_t^2/m_W^2$
\cite{leadingF,leadingB,leadingD,leadingC} and by logarithms
$\ln(m_t^2/m_W^2)$ \cite{logsK,logsP}. 
In both cases factorization is invalidated and nontrivial additional
terms are obtained.  

These results were derived by expanding in the ``small'' parameter
$m_W^2/m_t^2$. In contrast, no such expansion parameter is available
for the $Z$ decay rate into light $u$, $d$, $s$, or $c$ quarks.  All
these quarks are effectively massless and both $Z$ and $W$ bosons
appear as virtual particles in the relevant vertex diagrams which have
to be evaluated at $q^2 = m_Z^2$. A similar problem arises also for
the non-enhanced contributions to the $b\bar b$ vertex.  However, the
$Z$ decay rate is dominated by the $u$, $d$, $s$, and $c$ channels and
the corresponding mixed terms will dominate in the total $Z$ decay
rate.   For this reason we concentrate on light quark final state.

The case of QCD combined with QED provides an illustrative example for
mixed QCD-electroweak corrections.  The factorization ansatz $\left( 1
+ Q_q^2{3\over 4} { \alpha\over \pi}\right)
\left(1+{\alpha_s\over\pi}\right)$ implies a mixed term $Q_q^2{3\over
4} { \alpha\over \pi}{\alpha_s\over\pi}$.  However, the proper
evaluation \cite{Kataev} leads to $-{1\over 3} Q_q^2{3\over 4}
{ \alpha\over \pi}{\alpha_s\over\pi}$, which differs in magnitude and
sign from the naive ansatz.

With this motivation in mind we have evaluated the mixed corrections
of order ${ \alpha_{\rm weak}\over \pi}{\alpha_s\over\pi}$ for
$\Gamma_u$ and $\Gamma_d$ which originate from irreducible vertex
diagrams.  Our approach is based on the observation that the relevant
rates can be calculated for the decay of a virtual $Z$ with mass
squared equal $q^2$, with $q^2$ alternatively far smaller or far
larger than $m_Z^2$, by computing a large number of terms in the
expansion in $q^2/m_Z^2$ or $m_Z^2/q^2$.  The results can then be
extrapolated even to $q^2=m_Z^2$.


The tree-level decay width of the $Z$ boson into light quarks is given
by
\begin{eqnarray}
\Gamma^{(0)}(Z\to q\bar q) &=&  {\alpha N_C M_Z\over 6} 
\left[ (g_{q}^+)^2  + (g_q^-)^2 \right]
\label{eq:tree}
\\
g^+_q & =& -{s_W\over c_W} Q_q
\nonumber \\
g^-_q & =& {1\over s_Wc_W}(I_{3q}-s_W^2 Q_q)
\nonumber 
\end{eqnarray}
where $s_W$ and $c_W$ are sine and cosine of the weak mixing angle, 
$I_{3q}$ and $Q_q$ denote the isospin and electric charge of the quark
$q$, and $N_C=3$ is the number of colors.

The Born level result (\ref{eq:tree}) receives both QCD and
electroweak corrections.  The  QCD corrections have been calculated up
to three loops and yield a correction factor
\begin{eqnarray}
\Gamma^{QCD} = \Gamma^{(0)} \left(1 + {\alpha_s\over \pi} + \ldots\right)
\label{eq:qcdonly}
\end{eqnarray}

\begin{figure}[h]
\hspace*{-30mm}
\begin{minipage}{16.cm}
\[
\mbox{
\hspace*{-10mm}
\begin{tabular}{ccc}
\psfig{figure=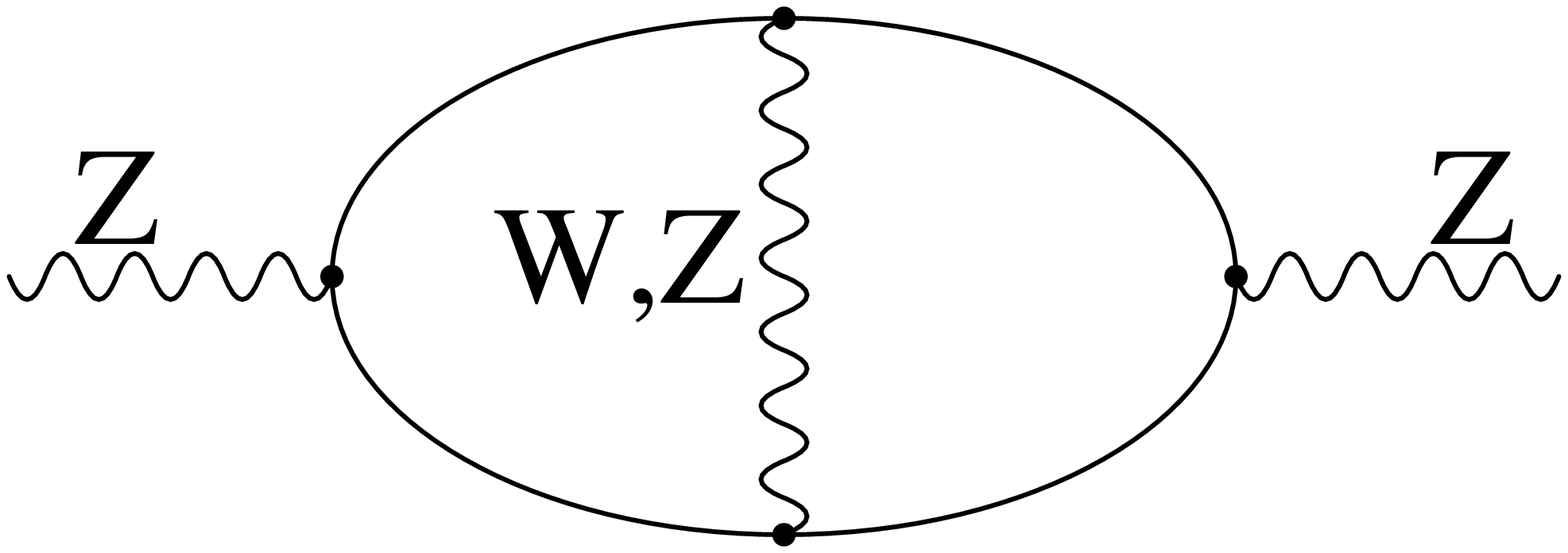,width=20mm,bbllx=210pt,%
bblly=410pt,bburx=630pt,bbury=550pt} 
&\hspace*{6mm}
\psfig{figure=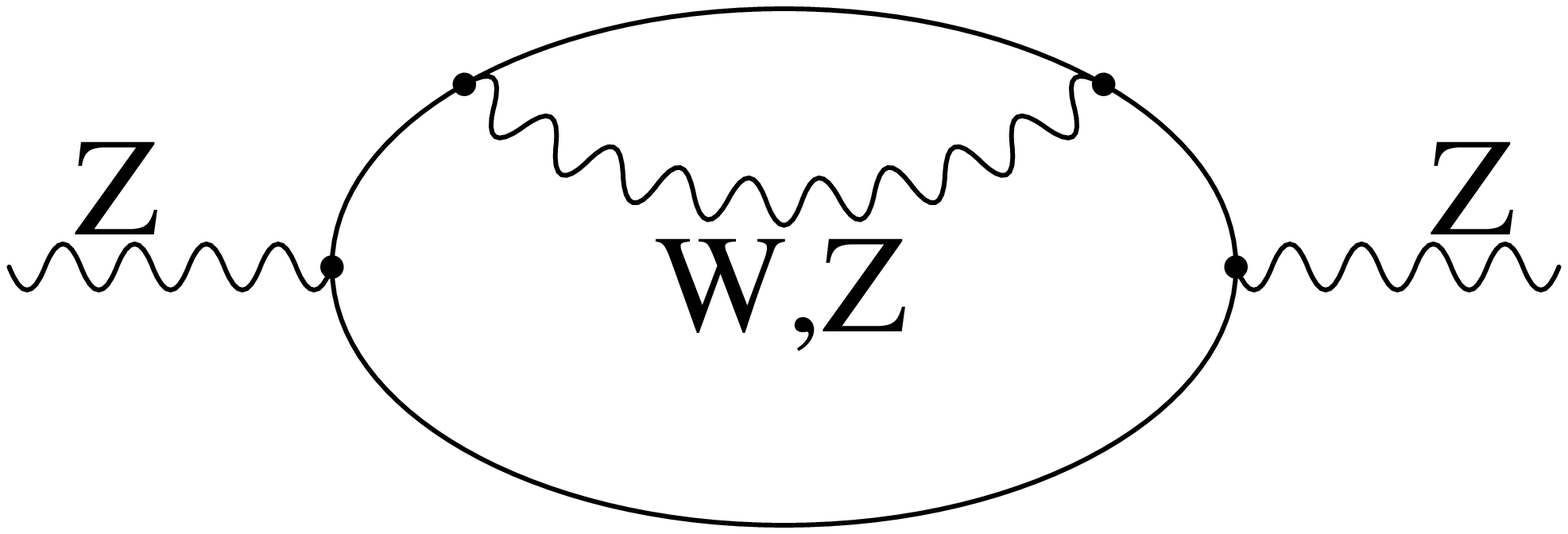,width=20mm,bbllx=210pt,%
bblly=410pt,bburx=630pt,bbury=550pt}
&\hspace*{6mm}
\psfig{figure=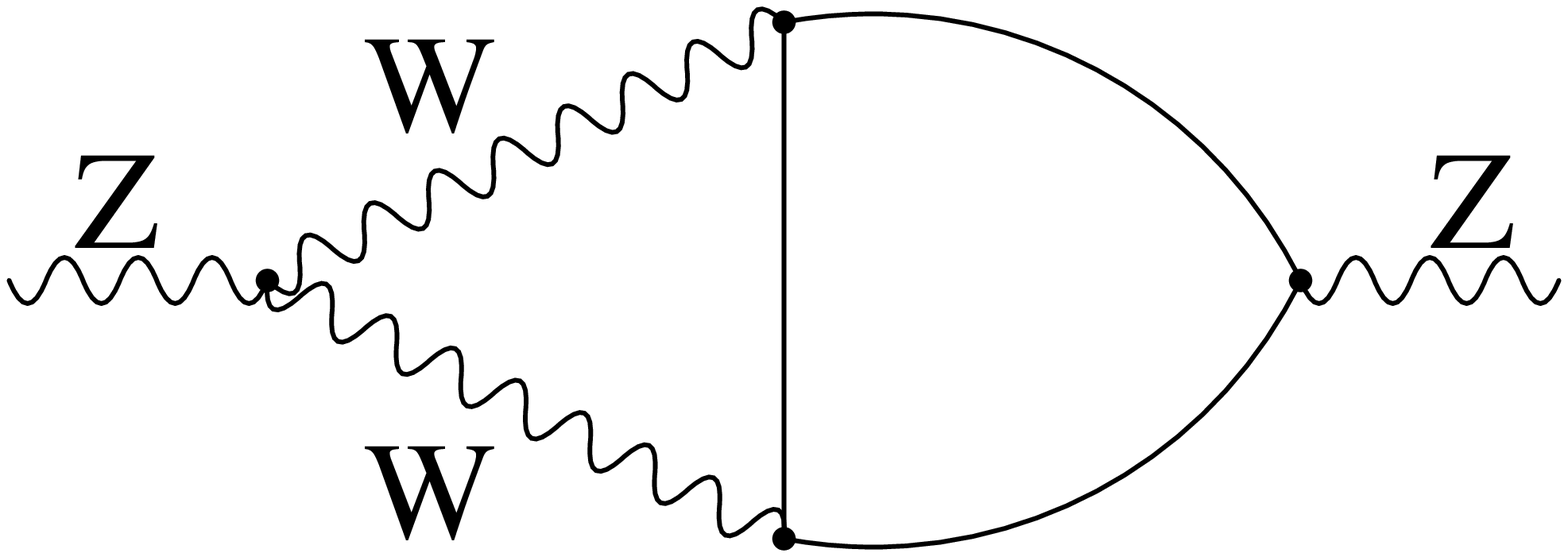,width=20mm,bbllx=210pt,%
bblly=410pt,bburx=630pt,bbury=550pt}
\\[10mm]
\rule{-17mm}{0mm} (a) &\hspace*{-.1cm} \rule{-4mm}{0mm}(b) & 
\hspace*{-.1cm}\rule{-2mm}{0mm}(c) 
\end{tabular}}
\]
\end{minipage}
\caption{One-loop electroweak corrections to the width of the $Z$ boson}
\label{fig:oneloop}
\end{figure}

First order electroweak corrections to the $Z$ boson hadronic width 
can be divided up into three finite parts 
\begin{eqnarray}
\Gamma^{\rm (1\ loop\ EW)}(Z\to q\bar q) 
\equiv \Gamma^{Z} +  \Gamma^{W} +  \Gamma^{ct} 
\end{eqnarray}
The first two contributions can be
calculated from the imaginary parts of the on-shell self-energy  
diagrams shown in fig.~(\ref{fig:oneloop}).  
$\Gamma^{Z}$ is given by the vertex correction (\ref{fig:oneloop}a)
with a $Z$ boson exchange
together with the $Z$ boson contribution to the wave
function renormalization of the quarks (\ref{fig:oneloop}b).
$\Gamma^{W}$ is given by two analogous diagrams with the internal $Z$
boson replaced by $W$, plus the diagram (\ref{fig:oneloop}c).  In the
linear 't~Hooft-Feynman gauge, adopted in this paper, the sum
of those three diagrams is ultraviolet divergent.  It is made
finite by including in $\Gamma^{W}$ the divergent part of the
counterterm generated by the $Z$ boson wave function renormalization;
this is obtained by making the following replacements in the $Zq\bar
q$ vertex:
\begin{eqnarray}
g^+_q\to 0, \qquad
g^-_q\to -{\alpha\over 4\pi} \left( {1\over \varepsilon} - \ln M_W^2\right)
{2c_W\over s_W^3} I_{3q}.
\label{eq:ct}
\end{eqnarray}
(The calculation is done using dimensional regularization in
$4-2\varepsilon$ dimensions.)

Finally, $\Gamma^{ct}$ is the tree-level decay width (\ref{eq:tree})
multiplied by the remaining, finite $Z$ boson wave function
renormalization constant.

The splitting of the decay rate into the three contributions is
convenient for the description of the QCD corrections.
In particular, the effect of QCD
corrections on $\Gamma^{ct}$ is just the multiplicative factor given
in eq.~(\ref{eq:qcdonly}).
For the remaining two contributions to the 
decay $Z \to q\bar{q}$ we have
\begin{eqnarray}
\Gamma^{Z}&=& {\alpha^2 N_C M_Z\over 12 \pi}
    \left[ (g_{q}^+)^4  + (g_q^-)^4 \right]
    \Lambda_2(x)
\nonumber\\
\Gamma^{W}&=& {\alpha^2 N_C M_Z\over 24 \pi s_W^2}
              g_q^-
\left[ 
g_{\tilde{q}}^- \Lambda_2(x)
+6 I_{3q} {c_W\over s_W} \Lambda_3(x)
\right]
\nonumber\\
x &=& {s\over M^2}
\end{eqnarray}
where $\tilde{q}$ is the isospin partner of the quark $q$, $s$ is the
momentum squared of the external $Z$ boson, and $M$ is the mass of the
virtual heavy boson inside the diagram.  For a decay of an on-shell
$Z$ boson we have $x=1$ in $\Gamma^Z$ and $x=M_Z^2/M_W^2\approx 1.292$
in $\Gamma^W$.
The functions $\Lambda_{2,3}$ can be expanded in a series in the
strong coupling constant
\begin{eqnarray}
\Lambda_i = \Lambda_i^{(0)} + C_F {\alpha_s\over \pi} \Lambda_i^{(1)} +
\ldots
\end{eqnarray}
where $C_F=4/3$ is the $SU(3)$ color factor.
The 0th order terms of these series 
have been calculated in \cite{Grz87,Bee88} 
\begin{eqnarray}
\Lambda_2^{(0)}(x) &=& -{7\over 2} - {2\over x}
 + \left( {2\over x} +3\right) \ln x
\nonumber \\ &&
 - 2\left( {1+x\over x} \right)^2
\left[ \ln x \ln ( 1 +x) + {\rm Li}_2 (-x) \right]
\nonumber\\
\Lambda_3^{(0)}(x) &=& {5\over 6} - {2\over 3x} 
 + {4+2x\over 3x}
 \sqrt{ {4\over x} -1} \arctan {1\over \sqrt{ {4\over x} -1}}
\nonumber \\ &&
 - {8\over 3} {2x+1\over x^2}\arctan^2 {1\over \sqrt{ {4\over x} -1}}
\end{eqnarray}
The behavior of these functions at $x\to 0$ and at $x\to
\infty$ can be found from asymptotic expansions for small and large
external momentum.  It turns out that only the first few terms
are needed to  compute with good accuracy 
$\Lambda_i^{(0)}$ 
for all values of the argument $x$.
This observation is the basis of the present paper.  Since
virtual gluons do not radically change the analytical properties of
the $Z$ self-energy diagrams we can approximate the QCD correction
functions $\Lambda_i^{(1)}$ by the first few terms of their asymptotic
expansion.  In a large part of the computations we employ the program
package MINCER \cite{MINCER} written in FORM \cite{FORM}.

\begin{figure}[h]
\hspace*{-40mm}
\begin{minipage}{16.cm}
\[
\mbox{
\hspace*{10mm}
\begin{tabular}{cc}
\psfig{figure=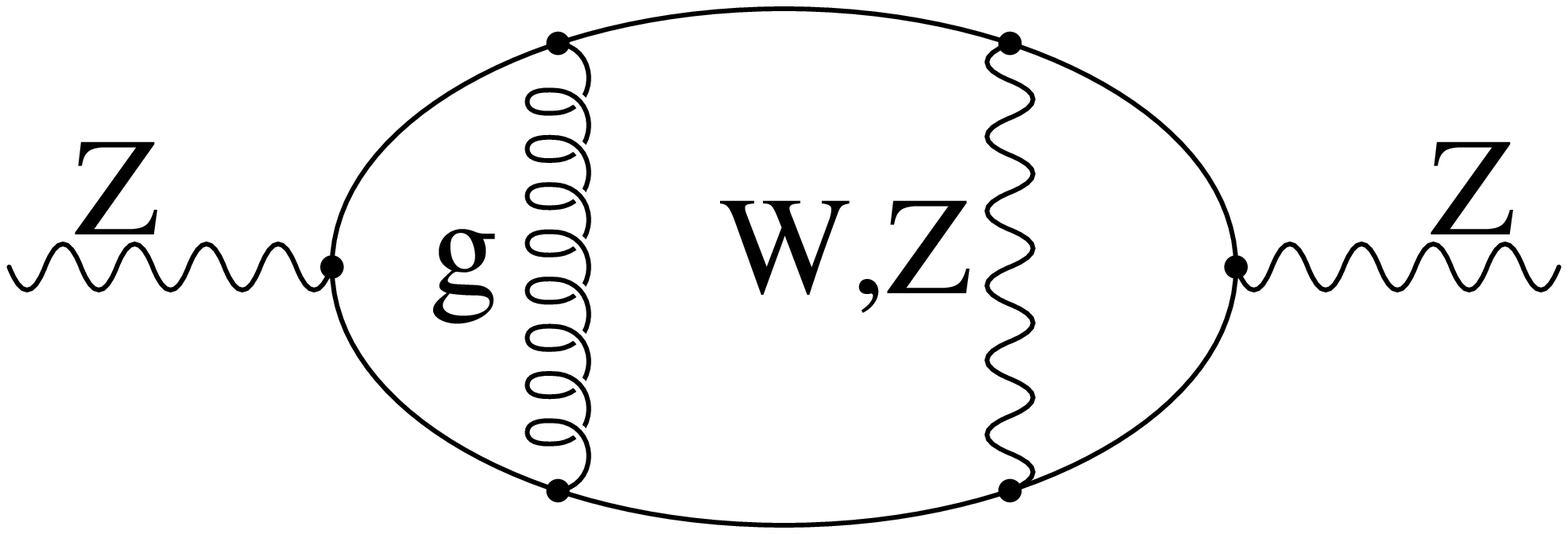,width=25mm,bbllx=210pt,%
bblly=410pt,bburx=630pt,bbury=550pt} 
&\hspace*{10mm}
\psfig{figure=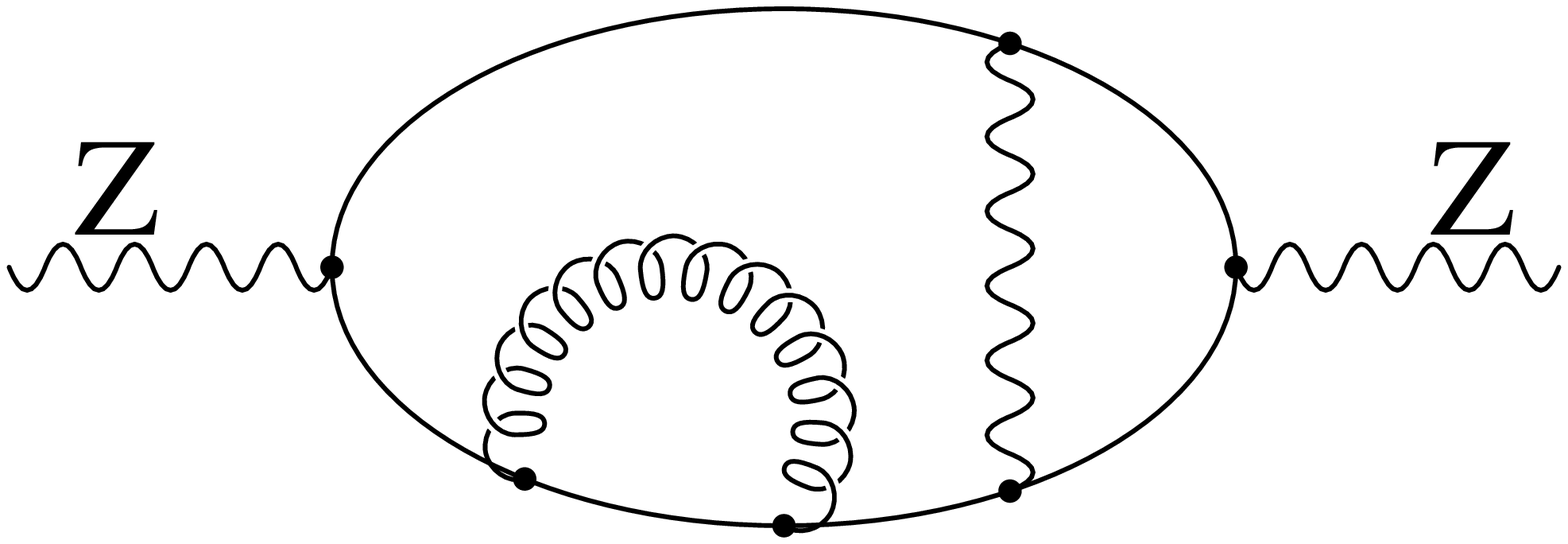,width=25mm,bbllx=210pt,%
bblly=410pt,bburx=630pt,bbury=550pt}
\\[8mm]
\hspace*{-15mm} (a) &\hspace*{-6mm} \rule{2mm}{0mm}(b) 
\\
\psfig{figure=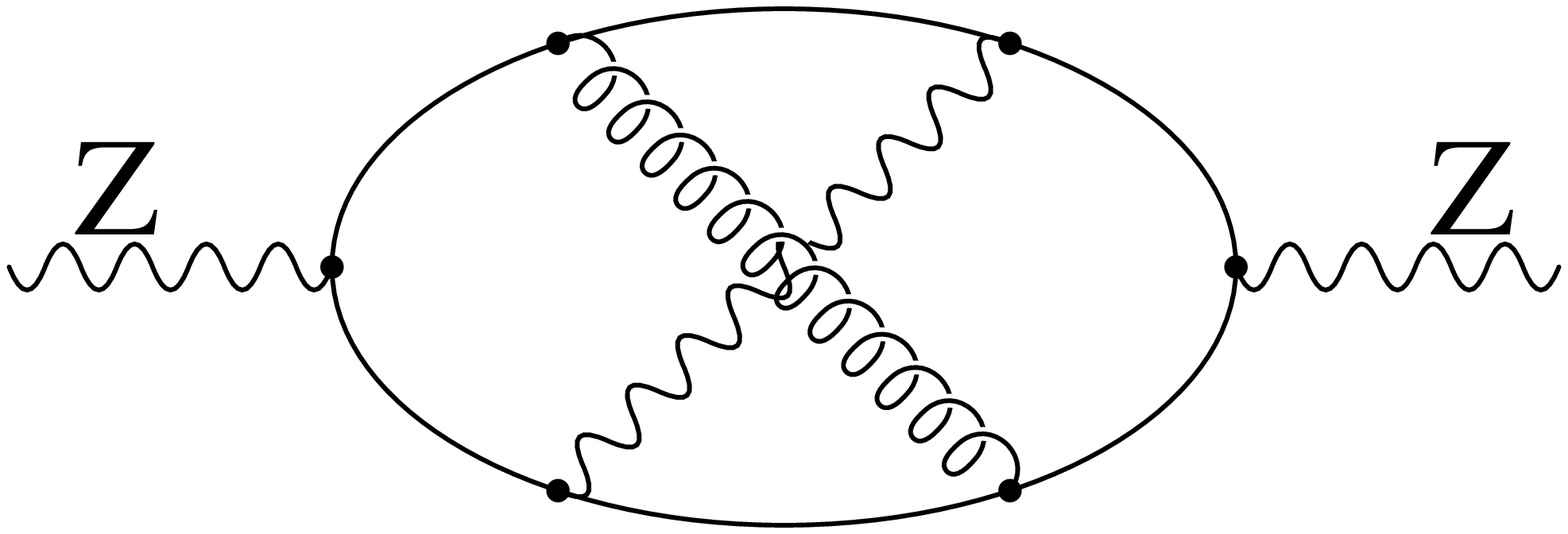,width=25mm,bbllx=210pt,%
bblly=410pt,bburx=630pt,bbury=550pt} 
&\hspace*{10mm}
\psfig{figure=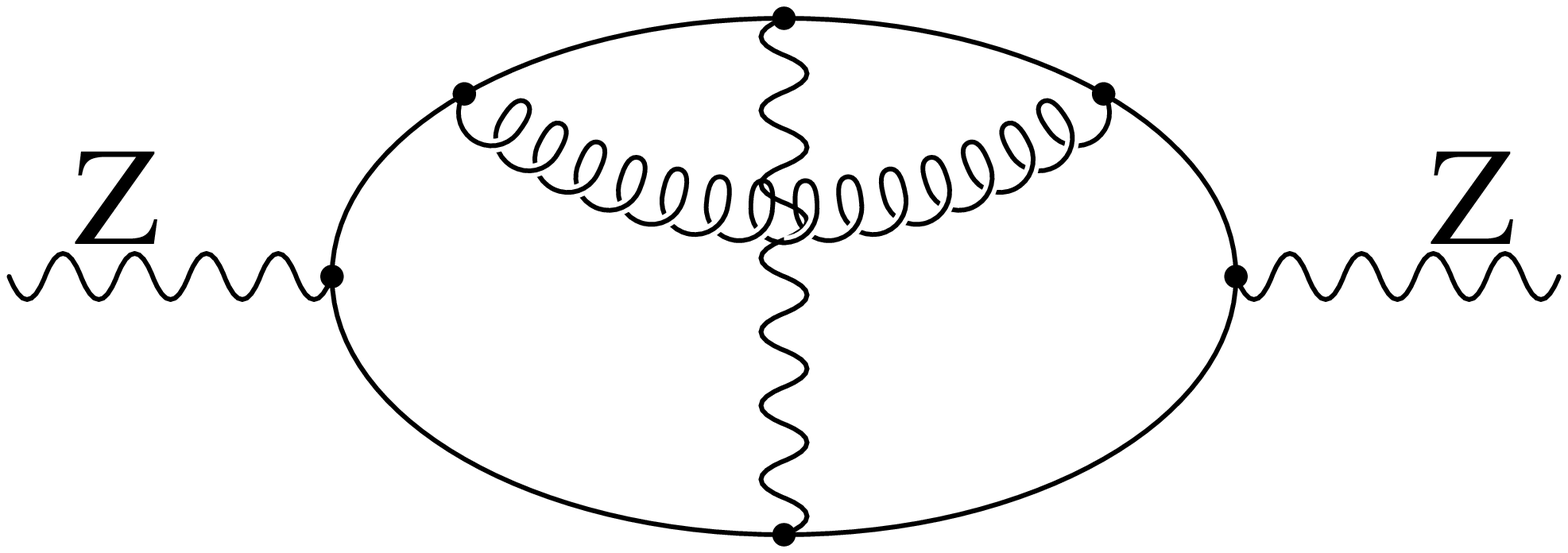,width=25mm,bbllx=210pt,%
bblly=410pt,bburx=630pt,bbury=550pt}
\\[8mm]
\hspace*{-15mm} (c) &\hspace*{-6mm} \rule{2mm}{0mm}(d) 
\end{tabular}}
\]
\end{minipage}

\caption{QCD corrections to the diagram 1(a)}
\label{fig:twoloopA}
\end{figure}

\begin{figure}
\hspace*{-40mm}
\begin{minipage}{16.cm}
\[
\mbox{
\hspace*{10mm}
\begin{tabular}{ccc}
\psfig{figure=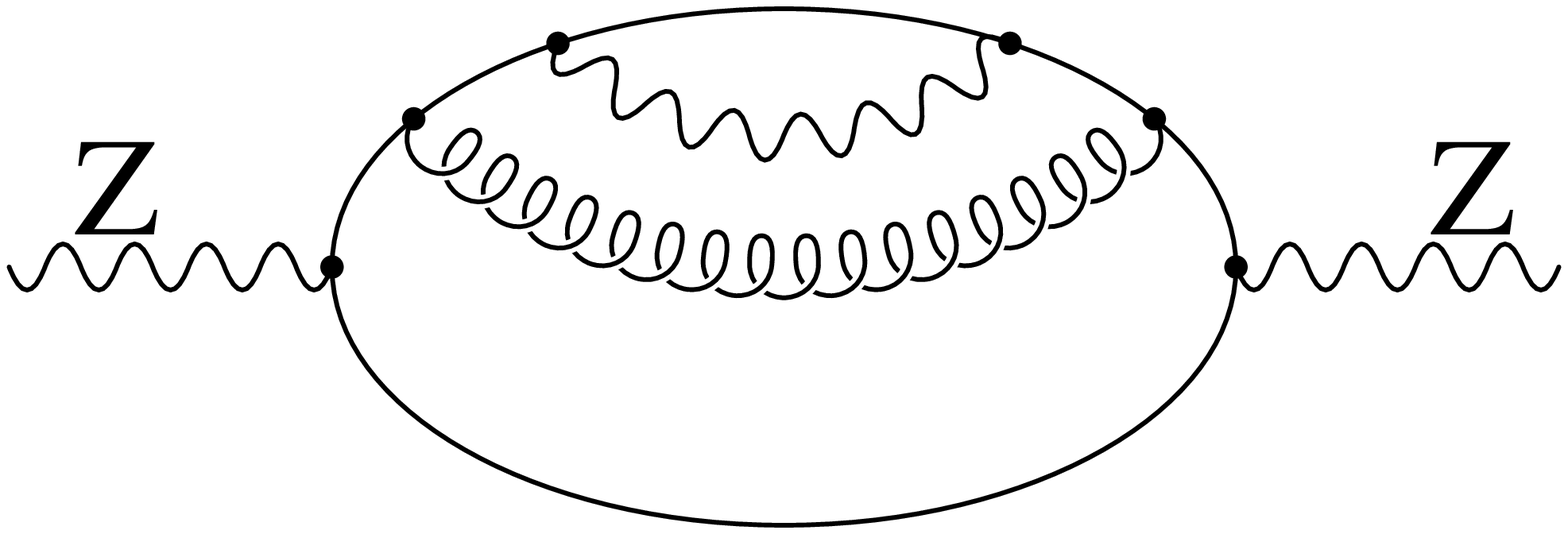,width=20mm,bbllx=210pt,bblly=410pt,%
bburx=630pt,bbury=550pt} 
&\hspace*{6mm}
\psfig{figure=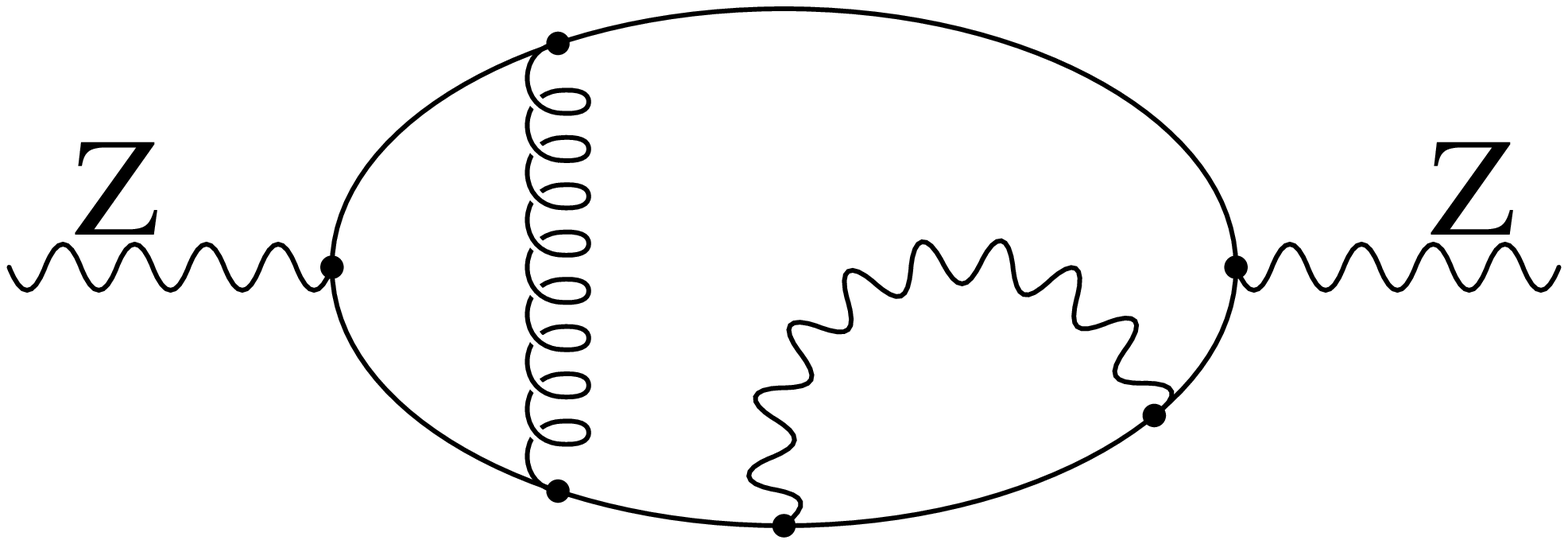,width=20mm,bbllx=210pt,bblly=410pt,%
bburx=630pt,bbury=550pt}
&\hspace*{6mm}
\psfig{figure=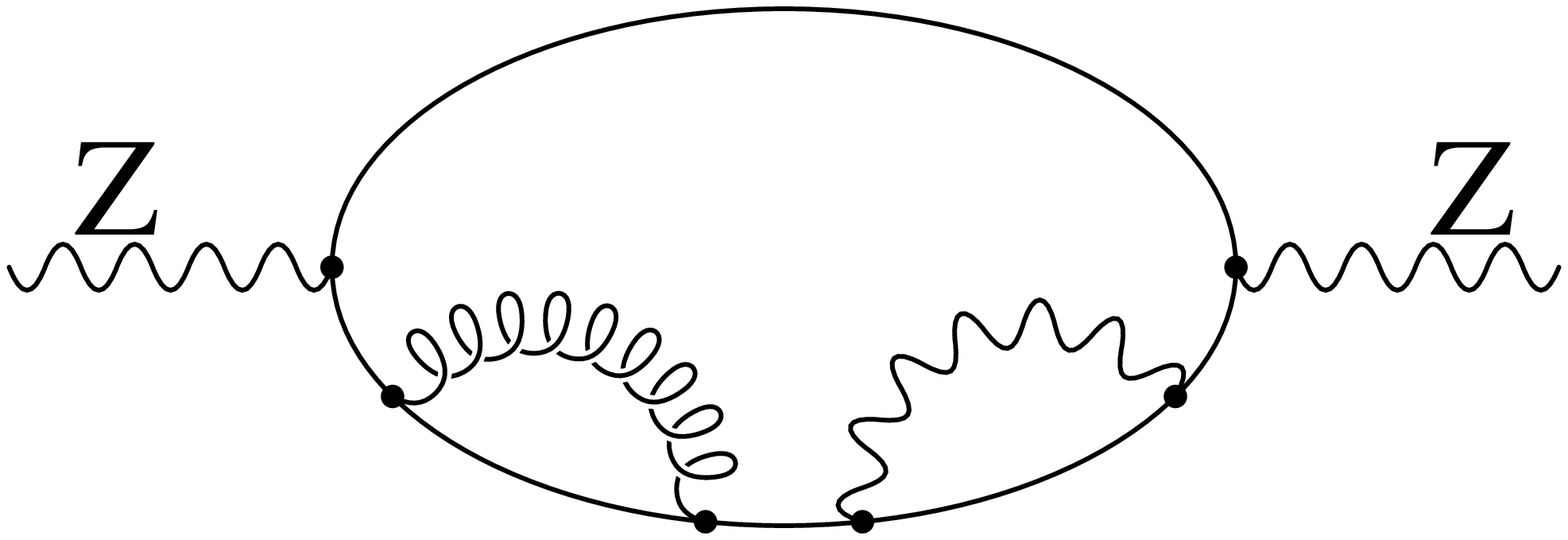,width=20mm,bbllx=210pt,bblly=410pt,%
bburx=630pt,bbury=550pt} 
\\[6mm]
\hspace*{-12mm} (a) &\hspace*{-6mm} (b) &\hspace*{-6mm} (c) 
\\
\psfig{figure=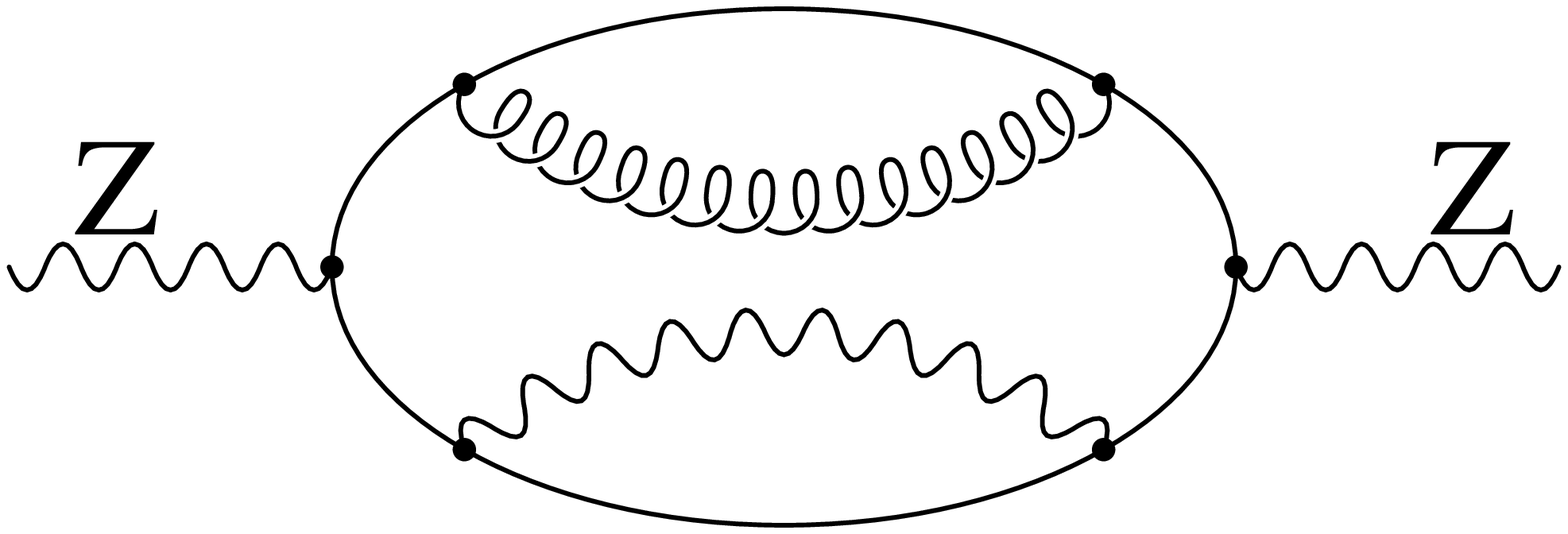,width=20mm,bbllx=210pt,bblly=410pt,%
bburx=630pt,bbury=550pt} 
&\hspace*{6mm}
\psfig{figure=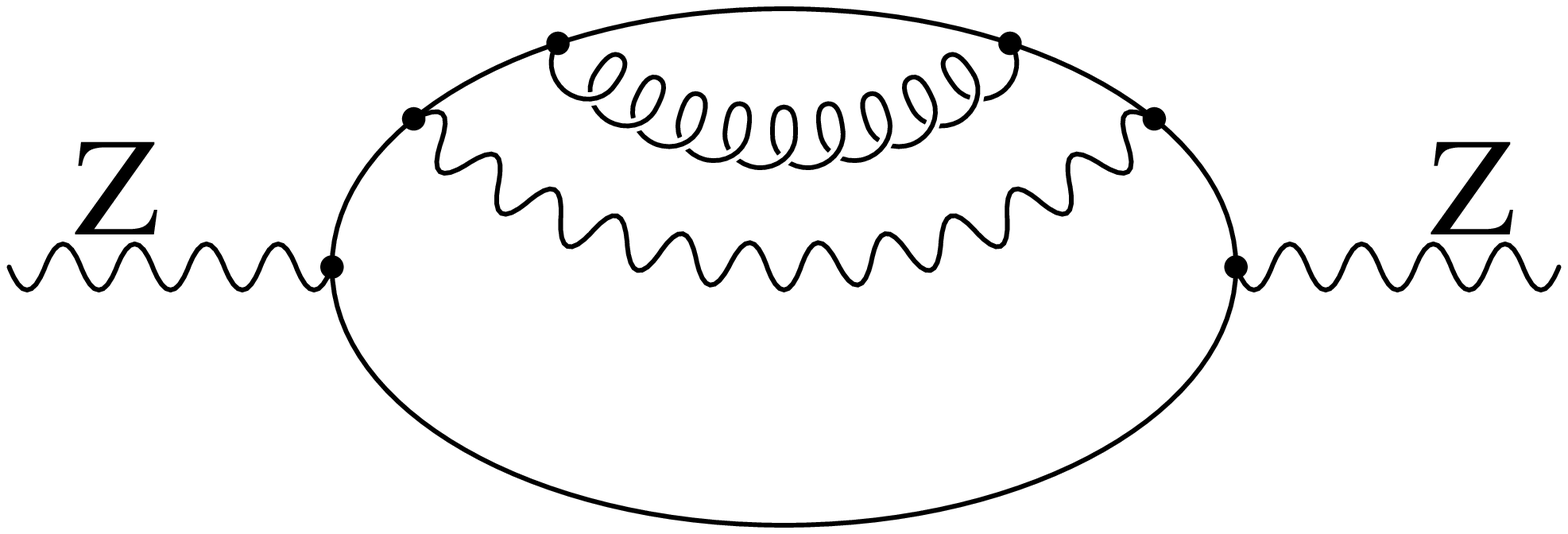,width=20mm,bbllx=210pt,bblly=410pt,%
bburx=630pt,bbury=550pt}
&\hspace*{6mm}
\psfig{figure=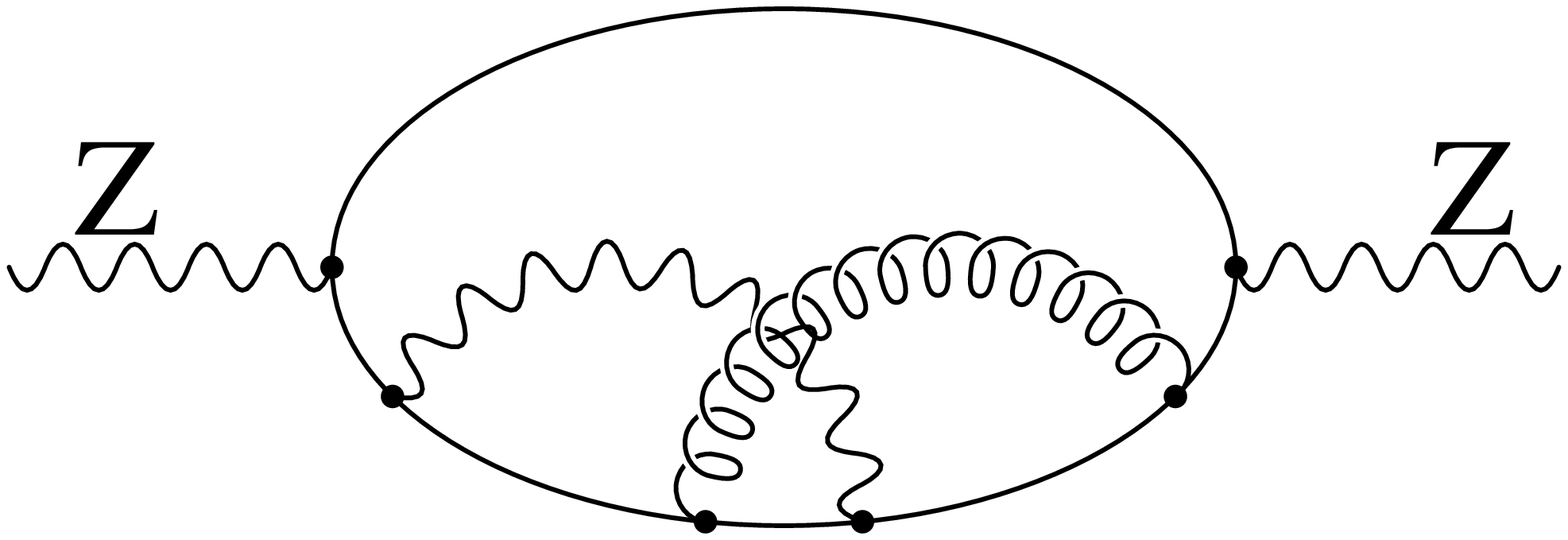,width=20mm,bbllx=210pt,bblly=410pt,%
bburx=630pt,bbury=550pt} 
\\[6mm]
\hspace*{-12mm} (d) &\hspace*{-6mm} (e) &\hspace*{-6mm} (f) 
\\
&\hspace*{6mm}
\psfig{figure=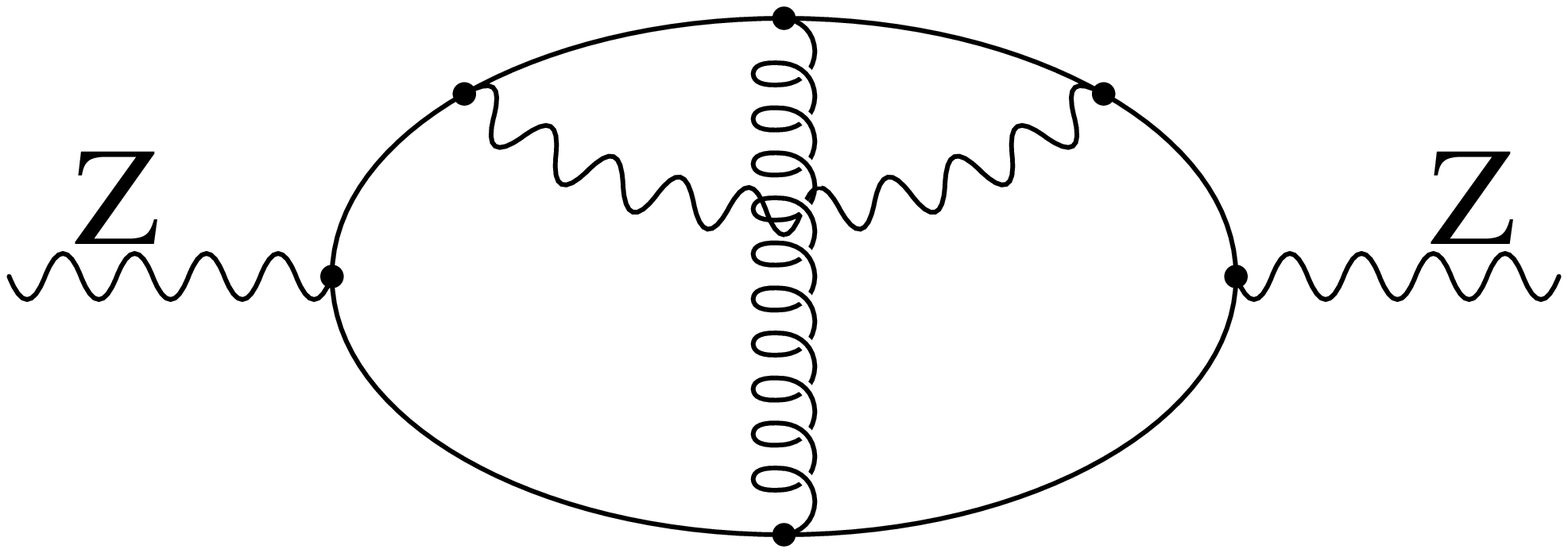,width=20mm,bbllx=210pt,bblly=410pt,%
bburx=630pt,bbury=550pt}
&
\\[6mm]
&\hspace*{-6mm} (g) &
\end{tabular}}
\]
\end{minipage}

\caption{QCD corrections to the diagram 1(b)}
\label{fig:twoloopB}
\end{figure}

\begin{figure}[h]
\hspace*{-40mm}
\begin{minipage}{16.cm}
\[
\mbox{
\hspace*{10mm}
\begin{tabular}{cc}
\psfig{figure=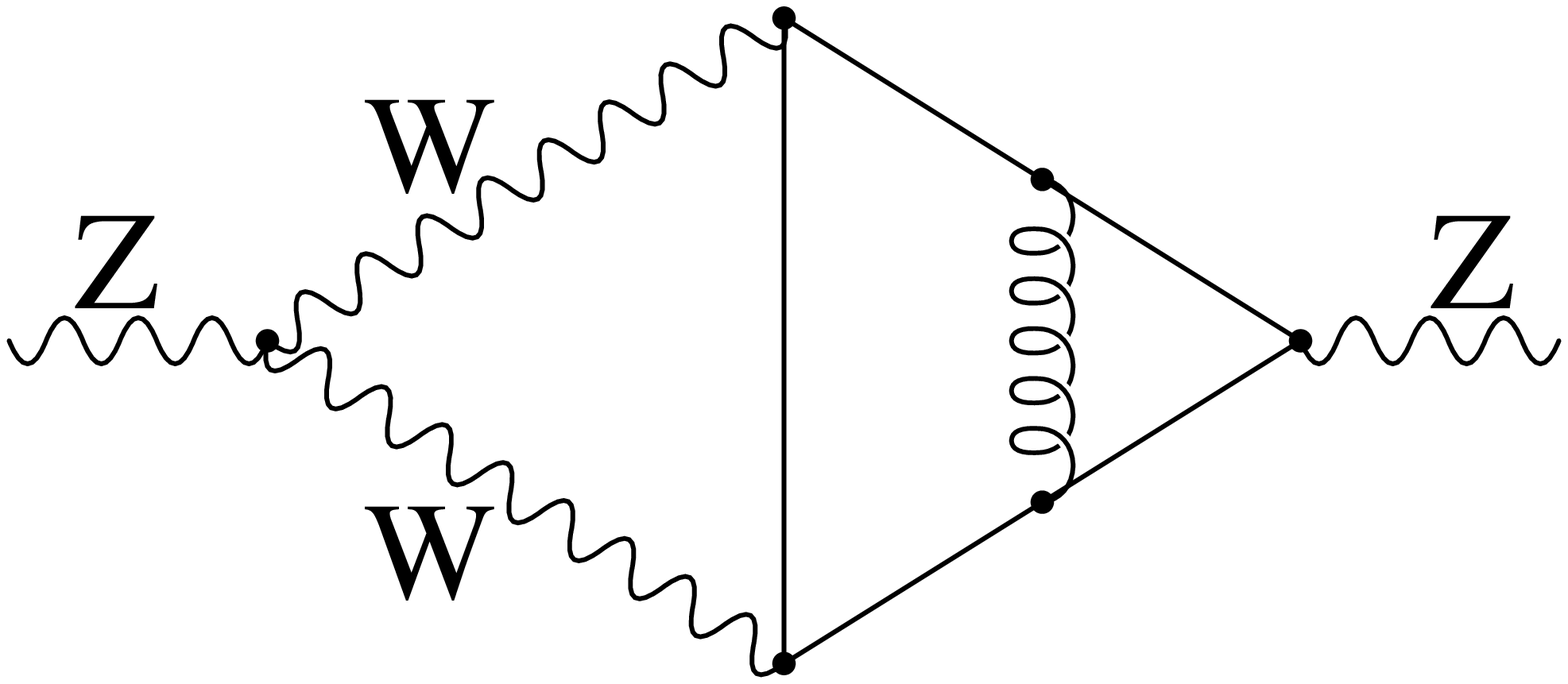,width=25mm,bbllx=210pt,bblly=410pt,%
bburx=630pt,bbury=550pt} 
&\hspace*{10mm}
\psfig{figure=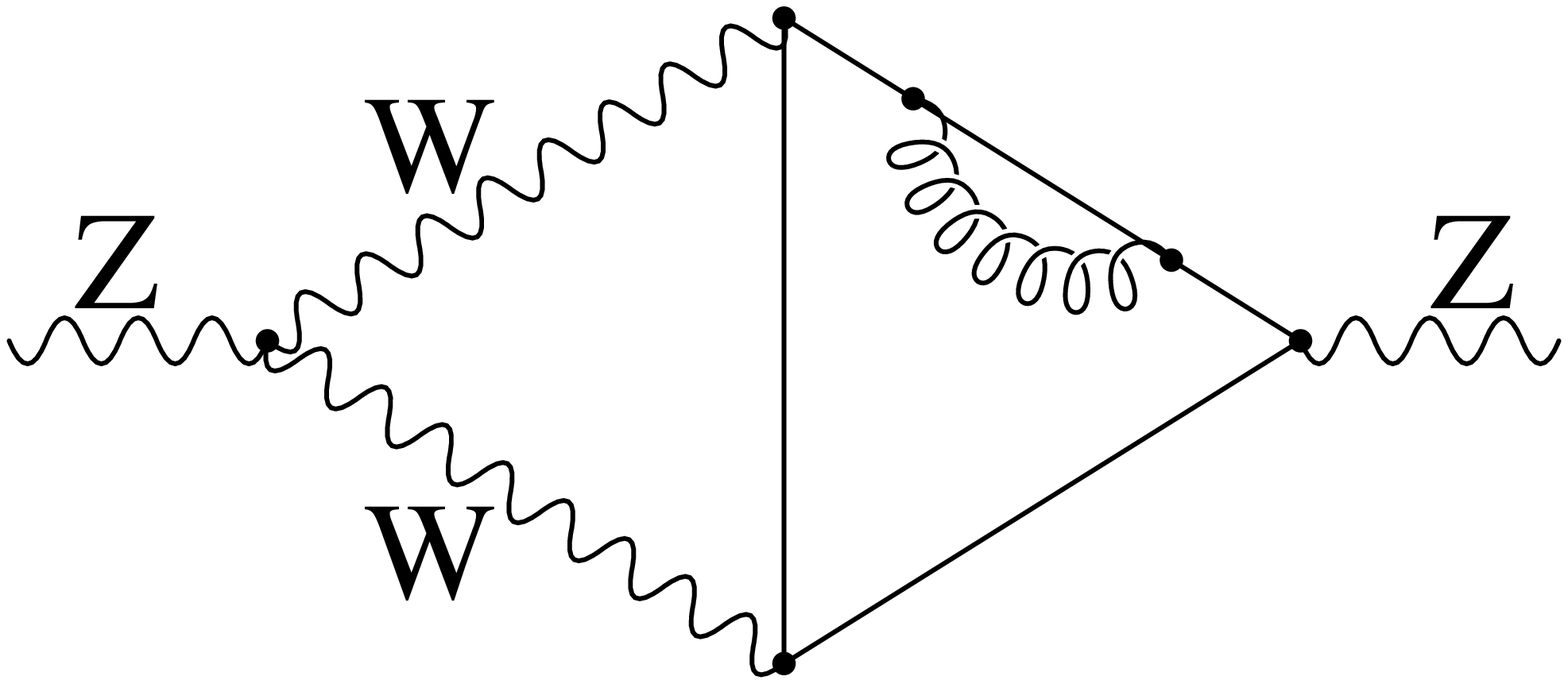,width=25mm,bbllx=210pt,bblly=410pt,%
bburx=630pt,bbury=550pt}
\\[8mm]
\hspace*{-15mm} (a) &\hspace*{-6mm} \rule{1mm}{0mm}(b) 
\\
\psfig{figure=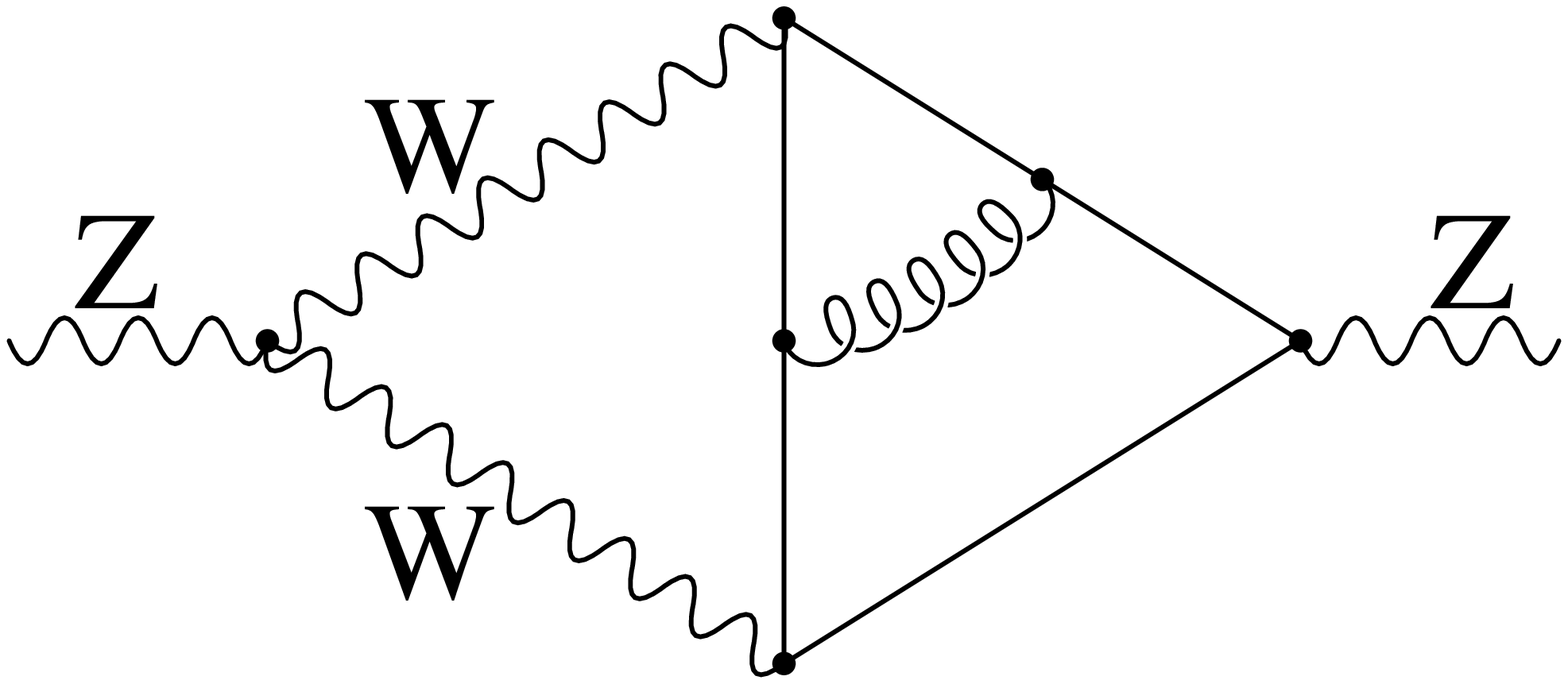,width=25mm,bbllx=210pt,bblly=410pt,%
bburx=630pt,bbury=550pt} 
&\hspace*{10mm}
\psfig{figure=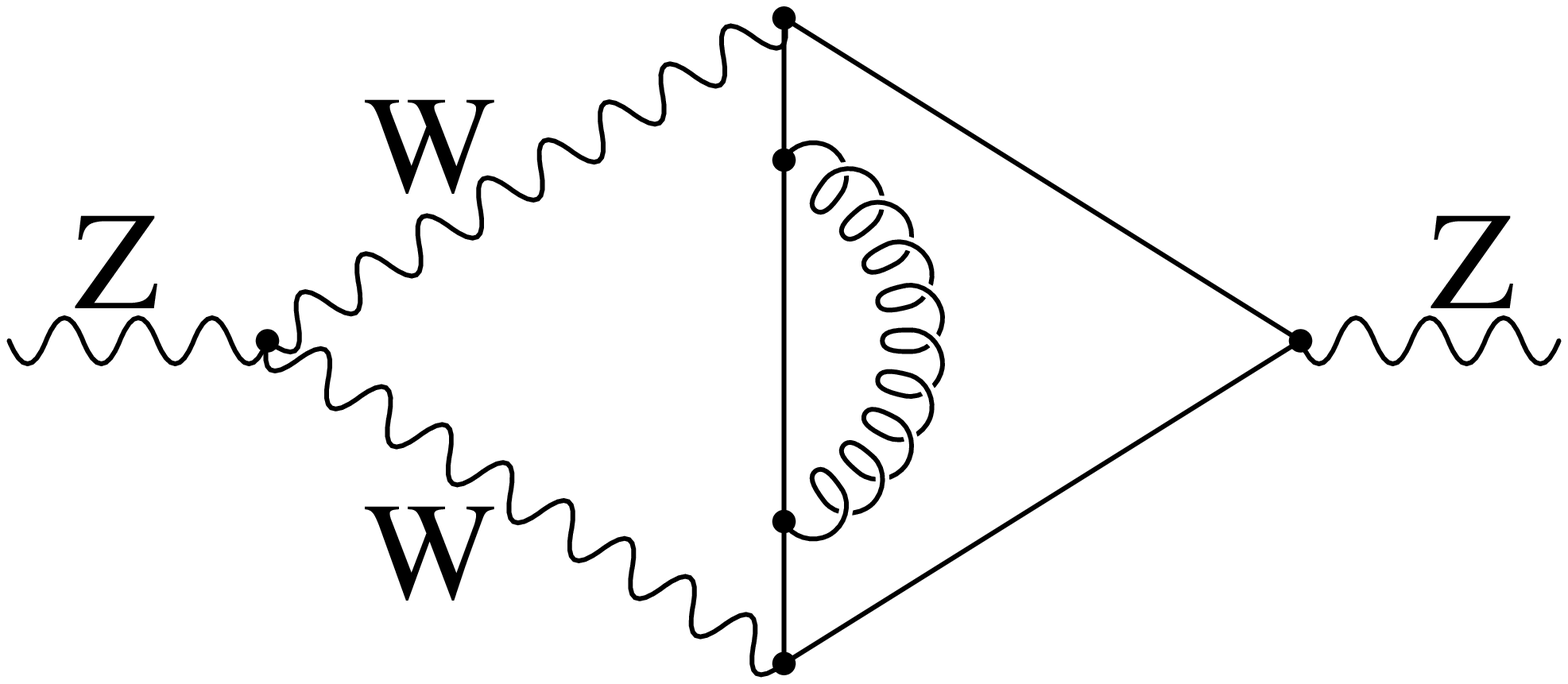,width=25mm,bbllx=210pt,bblly=410pt,%
bburx=630pt,bbury=550pt}
\\[8mm]
\hspace*{-15mm} (c) &\hspace*{-6mm} \rule{1mm}{0mm}(d) 
\end{tabular}}
\]
\end{minipage}

\caption{QCD corrections to the diagram 1(c)}
\label{fig:twoloopC}
\end{figure}

The function $\Lambda_2^{(1)}$ receives contributions from the
diagrams shown in fig.~(\ref{fig:twoloopA}) and
fig.~(\ref{fig:twoloopB}).  If the virtual particle is the $Z$ boson
the sum of these diagrams is finite.  On the other hand, in the case of
the virtual $W$ boson, the finite result is obtained only in the sum
with the diagrams of fig.~(\ref{fig:twoloopC}) and the counterterm
discussed before eq.~(\ref{eq:ct}).  This is a consequence of the
difference between the $Z$ couplings to the quarks in diagrams
\ref{fig:oneloop}a and \ref{fig:oneloop}b.

We notice, however, that the diagrams in fig.~(\ref{fig:twoloopB}) are
proportional to $(g^-_q)^2$ which can be rewritten as $g^-_q
g^-_{\tilde q} + 2I_{3q}g^-_q c_W/s_W$.  It is convenient to treat the
two
parts of this sum separately.  The first part together with
diagrams of fig.~(\ref{fig:twoloopA}) is finite and gives a contribution
described by $\Lambda_2^{(1)}$.  The second part can be combined with
diagrams in fig.~(\ref{fig:twoloopC}) and the counterterm to give
$\Lambda_3^{(1)}$.

We can write down an expansion of $\Lambda_2^{(1)}(x)$ around $x=0$ as
a sum of three series
\begin{eqnarray}
\Lambda_2^{(1)}(x) = S^S(x) + S^V_1(x) + S^V_2(x) \zeta(3)
\label{eq:DefLam2}
\end{eqnarray}
The first series is obtained from the finite part of the corrections
to the heavy boson contribution to the quark wave function
renormalization, fig.~(\ref{fig:twoloopB}).  The remaining two series
describe the gluonic effects on the heavy boson vertex correction
shown in fig.~(\ref{fig:twoloopA}); it is convenient to isolate the
terms containing $\zeta(3)$.  The results are

\begin{eqnarray}
\lefteqn{S^S(x)=
 \left[
 - {121\over 81}  
 - {1\over 9} \ln^2(x)
 + {22\over 27} \ln(x)
 \right] x}
\\ &&
+
 \left[
 - {169\over 5184}  
 - {1\over 144} \ln^2(x)
 + {13\over 432} \ln(x)
 \right] x^2
\nonumber \\ &&
+
 \left[
 - {2209\over 810000}  
 - {1\over 900} \ln^2(x)
 + {47\over 13500} \ln(x)
 \right] x^3
\nonumber \\ &&
+
 \left[
 - {1369\over 3240000}  
 - {1\over 3600} \ln^2(x)
 + {37\over 54000} \ln(x)
 \right] x^4
+{\cal O} (x^5)
\nonumber \\ 
\lefteqn{S^V_1(x) =
 \left[
 {71\over 18}  
 - {1\over 2} \ln^2(x)
 - \ln(x)
 \right] x}
\\ &&
+ \left[
 -{1159\over 1728}  
 - {7\over 24} \ln^2(x)
 + {5\over 4} \ln(x)
 \right] x^2
\nonumber \\ &&
+ \left[
 -{1853\over 7200}  
 - {13\over 60} \ln(x)
 \right] x^3
\nonumber \\ &&
+ \left[
 +{146179\over 648000}  
 - {11\over 120} \ln^2(x)
 + {83\over 540} \ln(x)
 \right] x^4
\nonumber \\ &&
+ \left[
 - {78941\over 396900}  
 - {37 \over 1890} \ln(x)
 \right] x^5
\nonumber \\ &&
+ \left[
 +{514328497 \over 3556224000 }  
 - {143\over 3360 } \ln^2(x)
 + {221 \over 8640 } \ln(x)
 \right] x^6
\nonumber \\ &&
+ \left[
 -{262758413 \over 2286144000 }  
 - {1\over 720} \ln^2(x)
 + {799 \over 90720} \ln(x)
 \right] x^7
\nonumber \\ &&
+ \left[
 +{13907067061 \over 160030080000}  
 - {403 \over 16800 } \ln^2(x)
 + {853 \over 4536000} \ln(x)
 \right] x^8
\nonumber \\ &&
+{\cal O} (x^9)
\nonumber \\ 
\lefteqn{S^V_2(x) = -2x+{1\over 2 } x^2 - {1\over 5} x^3 + {1\over 10}x^4}
\nonumber \\ &&
\qquad - {2\over 35}x^5
+ {1\over 28}x^6
- {1\over 42}x^7
+ {1\over 60}x^8
+{\cal O} (x^9)
\end{eqnarray}
In $S_1^V$ and $S^S$ we have not displayed terms
${\cal O}(x^0)$ which are divergent but 
cancel in the sum.  

$S^S$ and $S^V_2$ converge rapidly; in fact 
one can recognize the general formula for their coefficients and
sum up both series exactly; the result (being a rather complicated
function containing tetralogarithms) will be presented elsewhere.

On the other hand, 
$S_1^V$ converges very slowly for $x>1$; this is a consequence of a
three-particle cut at $s = M_Z^2$ in the diagrams in
fig.~(\ref{fig:twoloopA}).  This cut corresponds to the decay channel
$Z\to W^+d\bar u$.  If the internal heavy particle is a
$W$ boson we need the value of this series at $x=1.292$.  It is
therefore necessary to compute the expansion of the function
$\Lambda_2^{(1)}$ on the other side of the three-particle cut, that is
the asymptotic behavior at $x\to \infty$.  In analogy with
eq.~(\ref{eq:DefLam2}) we define
\begin{eqnarray}
\Lambda_2^{(1)}(x) = T^S(1/x) + T^V_1(1/x) + T^V_2(1/x) \zeta(3)
\label{eq:DefLamInfty}
\end{eqnarray}
In the sum $T^S+T^V_1$ the divergences cancel and the first three
terms give 
\begin{eqnarray}
-{3\over 8}\left(1-{1\over x^2}\right).
\label{eq:First3terms}
\end{eqnarray}
We do not present here the full formulas for the $T$-series for the
following reason:  beginning with the term $1/x^4$,
the coefficients of $1/x^n$ are identical (up to the sign of
the logs) to the
coefficients of $x^{n-2}$ in the corresponding series $S$.  Also in
the terms $1/x^3$ and $x$ there is an equality in the sums $T^S+T^V_1$
and $S^S+S^V_1$.  This remarkable feature gives us confidence in the
result.  It should be stressed that the $S$ and $T$ series result from
very different calculations.

The equality of the coefficients guarantees that both expansions
give equal results at $x=1$.  However, the slope of
both approximations is not quite the same at $x=1$; we can take the
magnitude of the resulting cusp as an estimate of the numerical error
in the final result.

We also notice that values of $\Lambda_2^{(1)}$ at
$x\approx 1$, relevant for the mixed QCD/electroweak corrections,  are
well approximated by the asymptotic value  $\Lambda_2^{(1)}(x\to
\infty)=-3/8$, which corresponds to the two loop mixed QCD/QED calculation
\cite{Kataev}.   We use in the numerical analysis 
$\Lambda_2^{(1)}(1) \approx \Lambda_2^{(1)}(1.292) \approx - 0.37\pm
0.04$. 

The calculation of $\Lambda_3^{(1)}$
involves four three-loop diagrams shown in
fig.~(\ref{fig:twoloopC}).  In the present analysis of the QCD
corrections we neglect the influence of the real
$W$ emission \cite{MarcWyl79,Braaten88,Glover89}.  (We notice
that a part of it, originating from $W$ emission off quarks, has
been included in the $T$ series discussed above.  In principle this partial
treatment is not gauge invariant; the induced error is, however,
negligible due to the phase space suppression of the $W$ emission.)

Taking into account only those cuts of diagrams in
fig.~(\ref{fig:twoloopC}) which do not cut $W$ lines we obtain
\begin{eqnarray}
\lefteqn{\Lambda_3^{(1)}(x)  = -{1\over 72}
\left( 14x + {89\over 60}x^2 
+ {116 \over 525}x^3 
+{53 \over 1400}x^4
\right. }\nonumber\\
&&
\left.
+{851 \over 121275}x^5
+{1381 \over  1009008}x^6
+ {\cal O}(x^7)
\right)
+{1\over 3x^2} S^S\left({1\over x}\right)
\nonumber \\
&&-{29\over 24}
+{3\over 8x}(1+2\ln x)
+{1\over 8x^2}(7-2\ln x)
-{1\over 24x^3}
\label{eq:ww}
\end{eqnarray}
The first part of this formula represents the finite part of the
diagrams in fig.~(\ref{fig:twoloopC}); the remaining terms arise from
the addition of the part of fig.~(\ref{fig:twoloopB}), as discussed
above eq.~(\ref{eq:DefLam2}).  We find $\Lambda_3^{(1)}(1.292) =
-0.87(1)$.


The mixed QCD/EW two-loop corrections are often naively
estimated in the factorized form, in which one assumes that this
effect is equal to the one-loop EW correction multiplied by the
one-loop QCD factor $\alpha_s/\pi$.  Our study shows, however, that the
assumption of factorization is misleading; for example the QCD
correction to the function $\Lambda_2$ has a relative minus sign.  

We can summarize our result as a difference between the full
corrections and the factorization formula
\begin{eqnarray}
\lefteqn{\Gamma^{\mbox{\small (2 loop EW/QCD)}} - {\alpha_s\over \pi}
\Gamma^{\mbox{\small (1 loop EW)}}}
\nonumber \\
&=& {\alpha_s\over \pi}
{\alpha^2N_C M_Z\over 12 \pi}
\left\{
    \left[ (g_{q}^+)^4  + (g_q^-)^4 \right]
    \left[ C_F\Lambda_2^{(1)}(1) - \Lambda_2^{(0)}(1) \right]
\right.
\nonumber \\
&&
+ {g_q^- \over 2 s_W^2}
\left[ 
g_{\tilde{q}}^- 
    \left[ C_F\Lambda_2^{(1)}(1.292) - \Lambda_2^{(0)}(1.292) \right]
\right.
\nonumber \\
&&
\qquad
\left.
\left.
+6 I_{3q} {c_W\over s_W} 
    \left[ C_F\Lambda_3^{(1)}(1.292) - \Lambda_3^{(0)}(1.292) \right]
\right]
\right\}
\label{eq:ins}
\end{eqnarray}
The ambiguity of the finite part of the counterterm (eq.~\ref{eq:ct})
cancels in this combination.
Inserting into eq.~(\ref{eq:ins}) the values of functions $\Lambda$ 
\begin{eqnarray}
\begin{array}{lclclcl}
\Lambda_2^{(0)}(1) & = & 1.080 & \qquad & 
\Lambda_2^{(1)}(1) & = & -0.37(4)
 \\
\Lambda_2^{(0)}(1.292) & = & 1.182 & \qquad & 
\Lambda_2^{(1)}(1.292) & = & -0.37(4)
 \\ 
\Lambda_3^{(0)}(1.292) & = & -0.288 & \qquad & 
\Lambda_3^{(1)}(1.292) & = & -0.87(1) 
\\ 
\end{array}
\nonumber
\end{eqnarray}
and using $\alpha_s=0.12$, $\alpha=1/129$,
$s_W^2=0.223$, $M_Z=91.19$ GeV,
we find that the net effect of the nonfactorizable corrections is
\begin{eqnarray}
\lefteqn{\Gamma^{\mbox{\small (2 loop EW/QCD)}} - {\alpha_s\over \pi}
\Gamma^{\mbox{\small (1 loop EW)}}}
\nonumber \\
&& \qquad\qquad = 
\left\{
\begin{array}{l}
-1.13(4)\times 10^{-4}  \mbox{ GeV}\quad \mbox{for $Z\to \bar uu$}
\\
 -1.60(6)\times 10^{-4}  \mbox{ GeV}\quad \mbox{for $Z\to \bar dd$}
\end{array}
\right.
\end{eqnarray}
The total change in the partial width $\Gamma(Z\to hadrons)$ is
obtained by summing over 2 down-type and 2 up-type quarks:
\begin{eqnarray}
\Delta \Gamma(Z\to u,d,s,c)= -0.55(3) \mbox{ MeV}
\end{eqnarray}
which translates into the change of the strong coupling constant
determined at LEP 1 equal to
\begin{eqnarray}
\Delta\alpha_s = 
-\pi{\Delta \Gamma(Z\to hadrons)\over \Gamma(Z\to hadrons) }
=\pi{0.55\over 1741} \approx 0.001
\end{eqnarray}
This shift is somewhat smaller but still of the same order of
magnitude as the 
experimental accuracy and should
to be taken into account in the  final analysis of LEP 1 data.

Acknowledgements:
A.C.~thanks Professor Wolfgang Hollik for a discussion on details of
ref.~\cite{Bee88} and pointing out ref.~\cite{Glover89}, and 
Kirill Melnikov and Matthias Steinhauser for helpful discussions and
advice. 
We thank R.~Harlander,  T.~Seidensticker, and M.~Steinhauser for
pointing out an error in eq.~(\ref{eq:ww}) in the first version of
this paper.
This research was supported by BMBF 057KA92P.


\begin{thebibliography}{10}

\bibitem{alpha3}
K. Chetyrkin, J. K{\"u}hn, and A. Kwiatkowski, {\em CERN Yellow Report 95-03}
  p.\ 175, and references therein.

\bibitem{ew}
D.~Bardin
 et~al., {\em CERN Yellow Report 95-03} p.\ 163, and references therein.

\bibitem{leadingF}
J. Fleischer, F. Jegerlehner, P. R{\c{a}}czka, and O.V.Tarasov, Phys. Lett.
  {\bf B293},  437  (1992).

\bibitem{leadingB}
G. Buchalla and A. Buras, Nucl. Phys. {\bf B398},  285  (1993).

\bibitem{leadingD}
G. Degrassi, Nucl. Phys. {\bf B407},  271  (1993).

\bibitem{leadingC}
K. Chetyrkin, A. Kwiatkowski, and M. Steinhauser, Mod. Phys. Lett. {\bf A8},
  2785  (1993).

\bibitem{logsK}
A. Kwiatkowski and M. Steinhauser, Phys. Lett. {\bf B344},  359  (1995).

\bibitem{logsP}
S. Peris and A. Santamaria, Nucl. Phys. {\bf B445},  252  (1995).

\bibitem{Kataev}
A. Kataev, Phys. Lett. {\bf B287},  209  (1992).

\bibitem{Grz87}
B. Grzadkowski, J.H. K\"uhn, P. Krawczyk, and R.G. Stuart,
Nucl. Phys. {\bf B281}, 18 (1987).

\bibitem{Bee88}
W. Beenakker and W. Hollik, Z. Phys. {\bf C40},  141  (1988).

\bibitem{MINCER}
S.~A. Larin, F.~V. Tkachov, and J. Vermaseren,
 preprint NIKHEF-H/91-18 (1991) (unpublished).

\bibitem{FORM}
J.~A.~M. Vermaseren, {\em Symbolic manipulation with {FORM}}, CAN, Amsterdam,
  1991.

\bibitem{MarcWyl79}
W. Marciano and D. Wyler, Z. Phys. {\bf C3},  181  (1979).

\bibitem{Braaten88}
E. Braaten and A. Kumar, Phys. Rev. {\bf D37},  3349  (1988).

\bibitem{Glover89}
E.~W.~N. Glover and J.~J. {van der Bij} (conv.),  
  in G. Altarelli, R. Kleiss, and C. Verzegnassi (eds.),
  {\em Z Physics at LEP 1},  CERN Yellow Report 89-08.

\end{thebibliography}


\end{document}